\documentclass[12pt]{iopart}

\usepackage[colorlinks=true,linkcolor=blue,citecolor=blue,citecolor=red]{hyperref}
\usepackage{cite}

\expandafter\let\csname equation*\endcsname\relax
\expandafter\let\csname endequation*\endcsname\relax
\usepackage{amsmath}

\usepackage{amssymb}
\usepackage{bbold}

\usepackage{graphicx}

\newcommand{\diff}[1]{\mathop{}\!\frac{\mathrm{d}#1}{\mathrm{d}t}}
\newcommand{\pdiff}[2]{\mathop{}\!\frac{\partial #1}{\partial #2}}
\newcommand{\ppdiff}[2]{\mathop{}\!\frac{\partial^2 #1}{\partial #2^2}}
\newcommand{\diffint}[1]{\mathop{}\!\mathrm{d}#1}

\begin{document}

\title[Stochastic resetting in a networked multiparticle system with correlated transitions]{Stochastic resetting in a networked multiparticle system with correlated transitions}

\author{Oriol Artime}

\address{CHuB Lab, Fondazione Bruno Kessler, Via Sommarive 18, 38123 Povo (TN), Italy}
\ead{oartime@fbk.eu}
\vspace{10pt}
\begin{indented}
\item[]\today
\end{indented}

\begin{abstract}
The state of many physical, biological and socio-technical systems evolves by combining smooth local transitions and abrupt resetting events to a set of reference values. The inclusion of the resetting mechanism not only provides the possibility of modeling a wide variety of realistic systems but also leads to interesting novel phenomenology not present in reset-free cases. However, most models where stochastic resetting is studied address the case of a finite number of uncorrelated variables, commonly a single one, such as the position of non-interacting random walkers. Here we overcome this limitation by framing the process of network growth with node deletion as a stochastic resetting problem where an arbitrarily large number of degrees of freedom are coupled and influence each other, both in the resetting and non-resetting (growth) events. We find the exact, full-time solution of the model, and several out-of-equilibrium properties are characterized as function of the growth and resetting rates, such as the emergence of a time-dependent percolation-like phase transition, and first-passage statistics. Coupled multiparticle systems subjected to resetting are a necessary generalization in the theory of stochastic resetting, and the model presented herein serves as an illustrative, natural and solvable example of such a generalization.
\end{abstract}

\section{Introduction}
The mechanism of randomly resetting a dynamical variable to a particular set of values is widespread in nature, such as in quantum mechanics, stochastic thermodynamics, chemical reactions and movement ecology~\cite{evans2020stochastic}. Stochastic resetting can be convenient for a number of reasons, and rooted in diverse origins, e.g., given by evolutionary advantage or designed to fulfill specific purposes. Such a mechanism can lead to the optimization of some task with respect to the reset-free scenario, e.g., in search strategies~\cite{kusmierz2014first}, enzymatic catalysis~\cite{rotbart2015michaelis}, rare event sampling~\cite{villen1991restart}, or Internet congestion reduction strategies~\cite{maurer2001restart}, among many others~\cite{evans2020stochastic}. From a fundamental viewpoint, in recent years the focus has been put on unveiling the properties of the relaxation to the stationary state, of the first-passage times and of the survival statistics in systems following diverse dynamics that undergo stochastic resetting~\cite{evans2011diffusion, evans2011diffusionb, mukherjee2018quantum, basu2019symmetric, grange2021aggregation, sandev2022heterogeneous}. Of particular note are the emergence of a nonequilibrium steady state and the existence of an optimal value of the resetting rate that minimizes the mean first-passage time (MFPT) to a target, properties that usually hold true for any embedding dimension~\cite{evans2014diffusion, riascos2020random, bressloff2021drift}, for several underlying dynamics followed by the particle~\cite{kusmierz2014first, evans2018run, masoliver2019telegraphic, gupta2019stochastic} and for most inter-resetting time statistics~\cite{pal2017first, chechkin2018random, nagar2016diffusion, kusmierz2019robust, masoliver2019anomalous}. See~\cite{evans2020stochastic} for a recent review.

Beyond particles diffusing in regular and heterogeneous topologies, resetting can be actually worth exploring in any dynamical system with a stochastically evolving state~\cite{gupta2014fluctuating, durang2014statistical, basu2019symmetric, grange2021aggregation}. An interesting example due to its theoretical and practical relevance is complex network growth undergoing stochastic resetting. Indeed, networks are abstract representations of complex systems, where nodes represent individual units and edges encode the interactions among these units~\cite{newman2003structure}, e.g., neurons and synapses in the brain, people and acquaintance ties in a social network, or web pages and hyperlinks in the world wide web. Apart from characterizing the topological structure of empirical networks~\cite{broder2000graph, colizza2006detecting, peixoto2014hierarchical, noldus2015assortativity, broido2019scale, voitalov2019scale, de2013mathematical, artime2022multilayer, battiston2020networks}, and study the behavior of dynamical models on top of them~\cite{arenas2008synchronization, pastor2015epidemic, porter2016dynamical, artime2017dynamics, hens2019spatiotemporal, artime2020abrupt}, it is crucial to understand how simple growing mechanisms yield the emergence of topological features present in real interconnected systems. Diverse mechanisms are able to predict or explain a plethora of these features: link rewiring leads to small-worldness~\cite{watts1998collective}, node addition with preferential attachment linking leads to scale-free networks~\cite{barabasi1999emergence, dorogovtsev2000structure, dorogovtsev2001effect, krapivsky2000connectivity}, or heterogeneous node fitness leads to \textit{winner-takes-all} (or Bose-Einstein condensation) phenomena~\cite{bianconi2001bose}.

Users in a social network can deactivate their accounts, websites in the WWW can be deleted, and neurons in the brain can be injured and stop functioning. In fact, many empirical networks might shrink, and even crash. Hence, a realistic ingredient to take into consideration in modeling network evolution is node removal~\cite{dorogovtsev2000scaling, sarshar2004scale, moore2006exact, srinivasan2007response, saavedra2008asymmetric, chung2004coupling, bauke2011topological, kalay2014fragmentation, crane2015cluster, zhang2016random}. It turns out that this effect can be framed as a general stochastic resetting problem that provides a conceptual advancement with respect to more traditional multiparticle stochastic resetting models, namely, the fact that the stochastic events on a particle influence the state of an arbitrary and variable number of other particles. Recently, there have been some contributions that take into account the particle interactions in the reset probability, but in an indirect way, see e.g.,~\cite{pelizzola2021simple, miron2021diffusion, miron2022local, bertin2022stochastic, krapivsky2022competition}. In this article we show that such a direct interaction can be mathematically handled and we derive some consequences from it: the emergence of a macroscopic network structure via a time-dependent phase transition, and the emergence of an inflection point in the mean first-passage time. 

The remainder of this paper is structured as follows. First, we explain in detail the model of network growth with node removal. We then write a master equation accounting for its probabilistic description, which is exactly solved. Afterward, the time-dependent percolation phase transition is examined and its critical point is derived. Finally, the analysis of the first-passage distribution and the mean first-passage time are considered. We close the article drawing some conclusions. 

\section{Model definition}

\begin{figure}[!t]
  \centering
  \includegraphics[width=0.8\columnwidth]{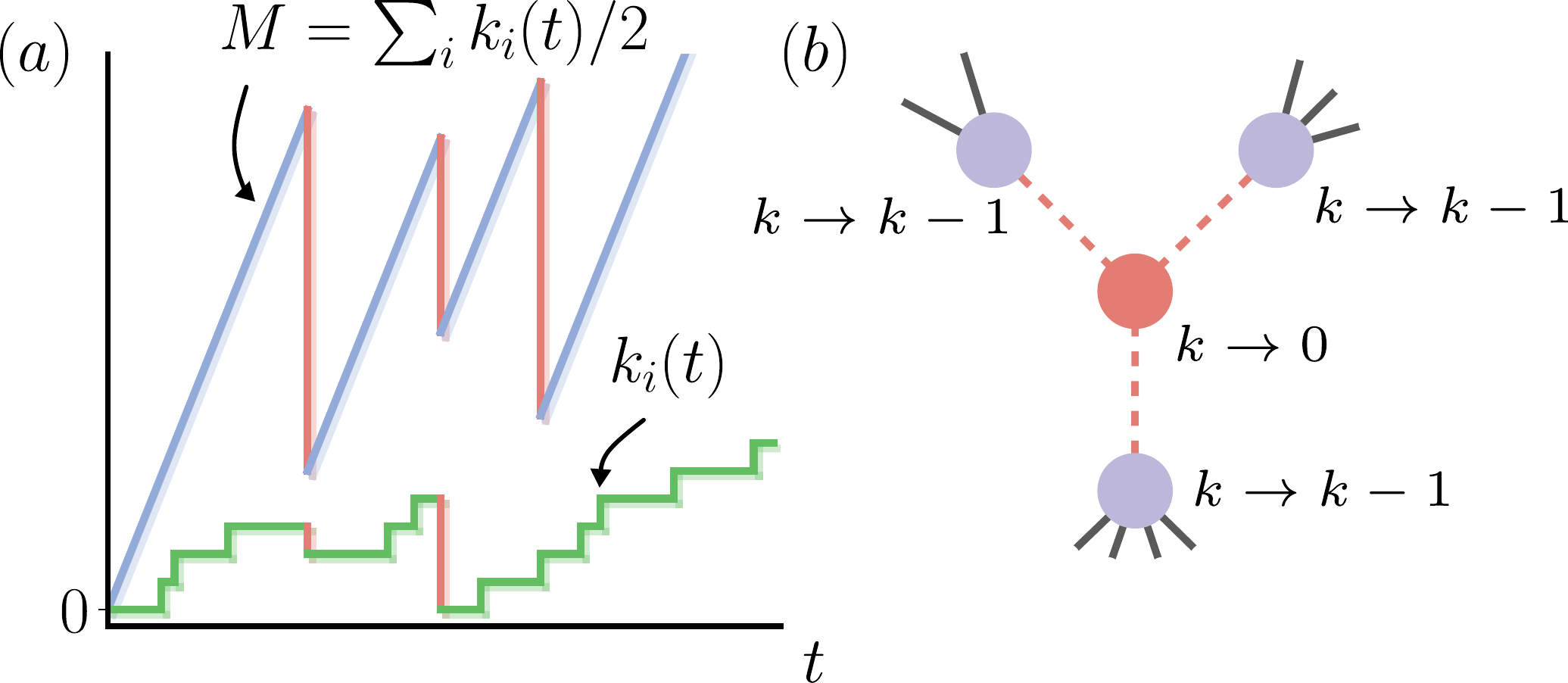}
\caption{In $(a)$, stylized time evolution of the number of links of node $i$, $k_i(t)$, and of the total number of links in the system, $M(t)$. The vertical drops in $M$ correspond to the resetting event of a node losing all of its links. In the first drop, one of $i$'s neighbor is reset, while in the second drop $i$ itself is reset. The last reset event does not affect on $i$ because it does not occur in its immediate vicinity. In $(b)$, resetting event on a node and the impact on its neighbors, which lose one degree unit.}
\label{fig:fig1}
\end{figure}

We consider a set of $N$ interconnected nodes, each one characterized by its degree $k$, i.e., the number of undirected connections to its neighbors. Two processes compete in the formation of the network. On the one hand, a link between two nodes picked uniformly at random is added at rate $\alpha N/2$. In the absence of resetting, the average degree would grow at rate $\alpha$, and if $N \to \infty$ this corresponds to a kinetic formulation of the Erd\H{o}s-R\'{e}nyi model with parameter (ratio between the number of links $M(t)$ at time $t$ and the number of possible links between $N$ pair of nodes) $p = \alpha t/N$, where $t$ is time~\cite{krapivsky2010kinetic}. On the other hand, with rate $rN$, a node selected uniformly at random is removed, along with all of its edges, and a new node with no connections is added into the network. Notice the coupling induced by the edges in both the growth and removal process: the increase of a degree unit occurs simultaneously in two nodes, and a deletion of a node with degree $k$ implies a degree unit loss in other $k$ nodes. See Fig.~\ref{fig:fig1} for a sketch of this process. With this model we prioritize the analytical understanding of the coupled multiparticle stochastic resetting process over a faithful description of network evolution. The incorporation of some realistic effects is possible ---e.g., the higher chances of well-connected nodes to acquire new links, the correlations between the degree and the removal probability, a variable system size $N$, or the fact that resetting event could make a node to lose not necessarily all its connections---, but they make the analytical treatment much more limited.

\section{Time-dependent degree distribution}

We can describe the temporal behavior of the model by studying the probabilities $p_k(t)$, the so-called degree distribution, that give the fraction of nodes with degree $k$ at time $t$~\cite{moore2006exact}. These probabilities follow the master equations
\begin{equation}
    \label{eq:MastEq}
    \diff{p_k} = \alpha p_{k-1} -  \alpha p_{k} - r p_{k} + r(k+1)p_{k+1} - r k p_{k} + \delta_{k,0} r.
\end{equation}
The first two terms are associated to the uniformly random addition of a link, while the rest correspond to the resetting process. In specific, the third term encodes the direct resetting of a node of degree $k$, while the fourth and the fifth terms correspond to the loss of one link due to the removal of a neighbor. The last term stands for the incoming flux of nodes that are reset to degree $0$. Negative degrees are not allowed, so the boundary condition is $p_{-1}(t) = 0$. As initial condition, we choose $p_k(0) = \delta_{k,0}$, that is, initially there are no links in the network. 

The exact full-time solution to the master equations~\eqref{eq:MastEq} can be obtained by means of the generating function. Introducing the time-dependent degree generating function $g(z,t) = \sum_{k=0}^{\infty} z^k p_k(t)$, if we multiply Eqs.~\eqref{eq:MastEq} by $z^k$ and sum over all degrees, we obtain 
\begin{equation}
    \pdiff{g}{t} = \left[ \alpha z - (\alpha + r) \right] g + r (1 - z) \pdiff{g}{z} + r,
\end{equation}
with the conditions $g(1,t) = 1$, which derives from the normalization of the degree distribution, and $g(z,0) = 1$, which derives from the initial condition. Employing the method of characteristics, the solution reads
\begin{equation}
    \label{eq:genfunc_sol}
    g(z,t) = \frac{1}{1-z} \left[ e^{\alpha z / r} \mathcal{G}\left((1-z)e^{-rt}\right) + \frac{r}{\alpha} \right],
\end{equation}
where $\mathcal{G}(x) = (x - r/\alpha) \exp \left[ \alpha/r \, (x-1) \right] $. To obtain the time-dependent degree distribution, we need to rewrite Eq.~\eqref{eq:genfunc_sol} as a power series of the auxiliary variable $z$. To do so, we use the well-known expansions $(1-x)^{-1} = \sum_{k=0}^{\infty} x^k$ and $e^{x} = \sum_{k=0}^{\infty} x^k/k!$, along with the relation
\begin{equation}
    \sum_{k=0}^{\infty} z^k \sum_{m=0}^{\infty} \frac{(x z)^m}{m!} = e^x \sum_{k=0}^{\infty} z^k Q(k+1,x),
\end{equation}
where $Q(a, b) \equiv \gamma(a,b)/\Gamma(a)$ is the regularized Gamma function, with $\Gamma(a)$ the complete Gamma function and
\begin{equation}
    \gamma(a,b) = \int_{b}^{\infty} \mathrm{d}x\, x^{a-1} e^{-x}
\end{equation}
the upper incomplete Gamma function. After some simple algebra, we arrive at our desired solution
\begin{equation}
    \label{eq:fulltime_sol}
    p_k(t) = \frac{r}{\alpha} \left[ 1 - Q(k+1,\,c(t)) \right] + e^{-c(t) - rt} \frac{c(t)^k}{k!},
\end{equation}
where we have introduced the function $c(t) = \frac{\alpha}{r} (1 - e^{-rt})$ to ease the notation.

\begin{figure}[!t]
  \centering
  \includegraphics[width=0.85\columnwidth]{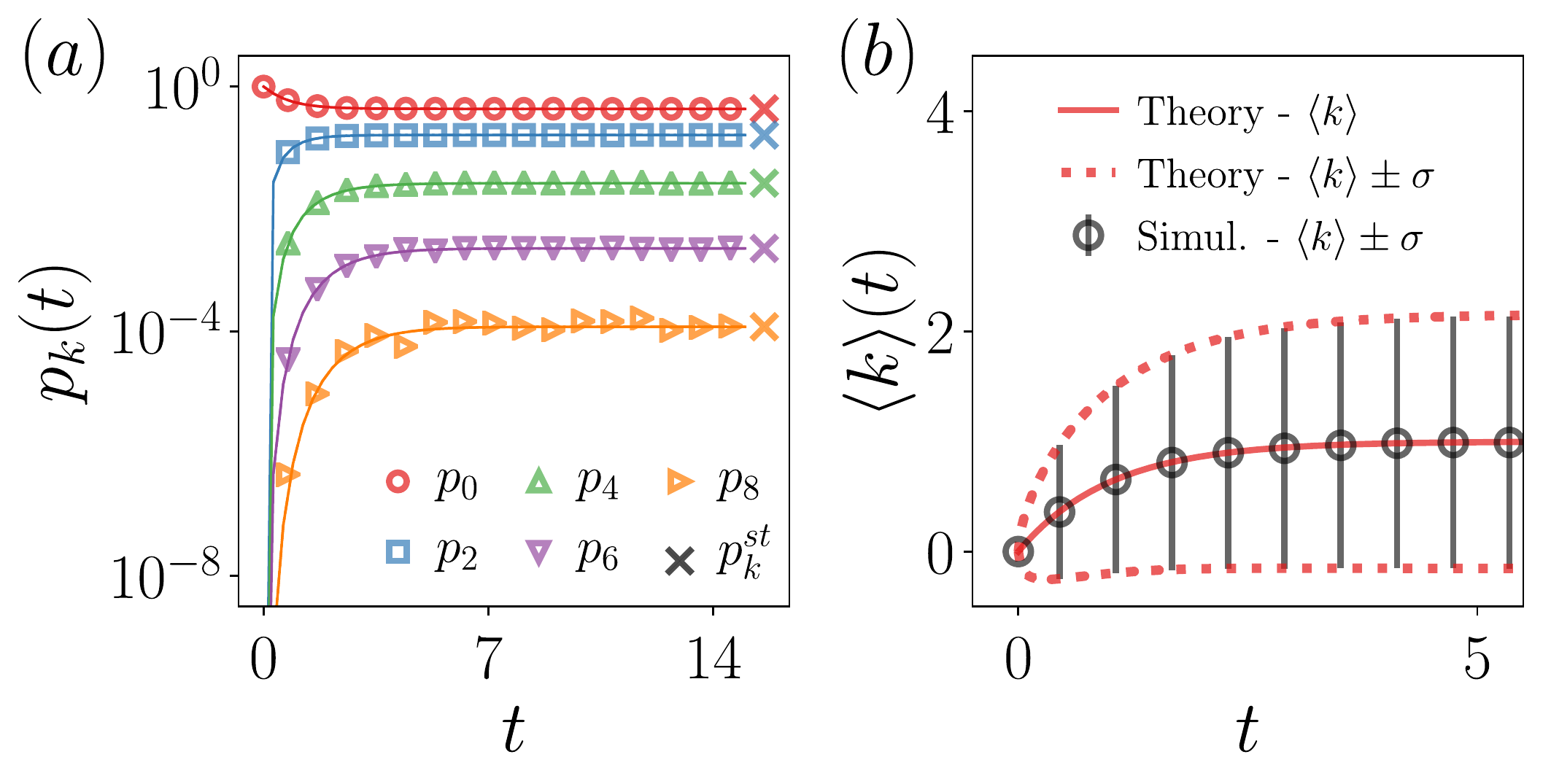}
\caption{In $(a)$, temporal evolution of the degree distribution. We show the curves for several degrees, indicated in the caption, together with their stationary value, computed from Eq.~\eqref{eq:StatDegDist}. Markers come from simulations and solid lines come from Eq.~\eqref{eq:fulltime_sol}. In $(b)$, temporal evolution of the mean degree and its variance. Markers come from simulations and lines from the theory. In all simulations we use $N = 1000$, $\alpha = 1$, $r=0.5$ and averages are computed over $200$ independent realizations.}
\label{fig:fig2}
\end{figure}

Although the full-time solution~\eqref{eq:fulltime_sol} is a complicated expression, we can extract some useful information under certain limits. Let us consider first the long-time limit, where we have the stationary degree distribution
\begin{equation}
    \label{eq:StatDegDist}
    p_k^{\text{st}} = \frac{r}{\alpha} \left[ 1 - Q \left(k+1, \frac{\alpha}{r}\right) \right].
\end{equation}
By virtue of the relation
\begin{equation}
    1 - Q \left(k+1, \frac{\alpha}{r}\right) = e^{-\alpha / r} \sum_{m=k+1}^{\infty} \frac{\left( \alpha / r \right)^m}{m!},
\end{equation}
which can be proved by integration by parts, Eq.~\eqref{eq:StatDegDist} becomes
\begin{equation}
    p_k^{\text{st}} = \frac{r}{\alpha} e^{-\alpha / r} \sum_{m=k+1}^{\infty} \frac{\left( \alpha / r \right)^m}{m!} \approx \left(\frac{\alpha}{r}\right)^k \frac{e^{-\alpha / r}}{(k+1)!},
\end{equation}
where in the last step we keep only the smallest value of the sum, which is at the same time the largest contribution. Using the Stirling's formula for the factorial, we obtain the asymptotic behavior of the stationary degree distribution
\begin{equation}
    p_k^{\text{st}} \approx e^{-\alpha / r} \left(\frac{e \, \alpha}{r}\right)^k (k+1)^{-3/2 - k},
\end{equation}
which decays much faster than an exponential~\cite{moore2006exact}. In Fig.~\ref{fig:fig2}$(a)$ we verify that both the analytical time-dependent degree distribution and its stationary values coincide with the values coming from the simulations of the model.

Another interesting limit to explore is the one of small reset rate $r$. On the one hand, if we set $r=0$ and solve the master equations~\eqref{eq:MastEq}, the time-dependent degree distribution is $p_k(t) = (\alpha t)^k e^{-\alpha t} / k!$. That is, in the reset-free scenario there is no stationary state, and the network grows indefinitely. On the other hand, if we take the limit of small resetting rates in the time-dependent solution~\eqref{eq:fulltime_sol}, we get
\begin{equation}
    \label{eq:degdist_expansion}
    \frac{(\alpha  t)^k e^{- \alpha t}}{k!} + r \left[ \frac{t (\alpha t -k - 2)}{2} \frac{(\alpha  t)^k e^{-\alpha t}}{ k!} + \frac{1-Q (k+1,t \alpha )}{\alpha }\right]+\mathcal{O}\left(r^2\right).
\end{equation}
We see that only when $r = 0$ Eq.~\eqref{eq:degdist_expansion} reduces to the reset-free degree distribution. As far as the resetting rate becomes finite, this no longer holds true and the system is able to tend to a well-defined stationary state. Thus, the introduction of a resetting rate, however small it may be, induces the emergence of a nonequilibrium steady-state.

Additional information of the process of network growth under stochastic resetting can be gained with the temporal evolution of the moments of the degree distributions $\langle k^n \rangle (t) \equiv \sum_k k^n p_k(t)$. Although they could be computed directly from the full-time expression~\eqref{eq:fulltime_sol}, a significantly easier approach is to solve the differential equation governing them. We can solve exactly up to arbitrary order $n$ because the differential equations for the moments are closed. For example, after some trivial algebra, for the mean degree we obtain
\begin{equation}
    \label{eq:meandeg_difeq}
    \diff{\langle k \rangle} = \alpha - 2r \langle k \rangle,
\end{equation}
with initial condition $\langle k \rangle (0) = 0$. The solution reads 
\begin{equation}
    \label{eq:meandeg}
    \langle k \rangle (t) = \frac{\alpha}{2 r} \left( 1 - e^{-2rt} \right),
\end{equation}
and it is displayed in Fig.~\ref{fig:fig2}$(b)$, along with the standard deviation, which can be trivially calculated. 

\section{Percolation transition}

The time-dependent degree distribution and their moments allow us to study network-wide connectivity properties during the growth process. As an illustration, we herein consider the emergence of the giant component, which is the order parameter of the percolation phase transition and is of crucial importance, because connectivity is usually assumed the first proxy for network functionality. Let us denote $u(t)$ the probability that, at time $t$, a node is not in the giant component via one of its links. If that node has $k$ connections, the probability to belong to the giant component is then $1 - u(t)^k$. Averaging over the degree distribution, we obtain the size of the giant component
\begin{equation}
    \label{eq:SizeGC}
    S(t) = \sum_{k=0}^{\infty} p_k(t) \left(1 - u(t)^k\right) = 1 - g(u,t),
\end{equation}
where $g(z,t)$ is the time-dependent degree generating function used to solve Eq.~\eqref{eq:MastEq}. To solve Eq.~\eqref{eq:SizeGC}, we need to know what is the value of $u(t)$. This can be obtained in a recursive manner, by noting that the condition for a node to not belong to the giant component
following one of its links is that the node at the end of the link we are following does not belong to the giant component via any of its other $k-1$ neighbors, i.e., $u(t)^{k-1}$. To account for the network heterogeneity, this latter quantity needs to be averaged in a similar fashion to the computation of the giant component $S(t)$, although without the use of the degree distribution. Indeed, let $q_{k}(t)$ denote the probability that a node at the end of a link has $k$ connections at time $t$. It then follows that $q_{k}(t)$ is proportional to $k p_{k}(t)$. The proportional factor must ensure the normalization of this distribution, that is, $q_{k}(t) = k p_{k}(t) / \langle k \rangle(t)$. Hence, we reach to our self-consistent equation for $u(t)$,
\begin{equation}
\label{eq:RecRel}
u(t) = \frac{1}{\langle k \rangle (t)} \sum_{k=0}^{\infty} k \, p_k(t)\, u(t)^{k-1} = \frac{\partial_u g(u,t)}{\partial_u g(1,t)}.
\end{equation}
Note that $u(t)=1$ is always a solution to Eq.~\eqref{eq:RecRel} and, if plugged in Eq.~\eqref{eq:SizeGC}, we obtain $S(t)=0$. This absence of the giant component is of course consistent with the physical meaning of $u(t)$, and it could have been guessed \textit{a priori}. When a second, non-trivial solution $u(t) \neq 1$ exists, we find a finite $S(t)$. The appearance of this second solution marks the existence of a percolation-like phase transition, and occurs when the derivatives of both sides of Eq.~\eqref{eq:RecRel}, evaluated at $u(t)=1$, are equal. In our case, the condition for criticality is given by the combination of values $(t,\alpha,r)$ that satisfy
\begin{equation}
    \label{eq:crit_point}
    1 = \frac{2\alpha}{3r} \frac{1-2e^{-rt} + e^{rt}}{1 + e^{rt}}.
\end{equation}

\begin{figure}[!t]
  \centering
  \includegraphics[width=0.9\columnwidth]{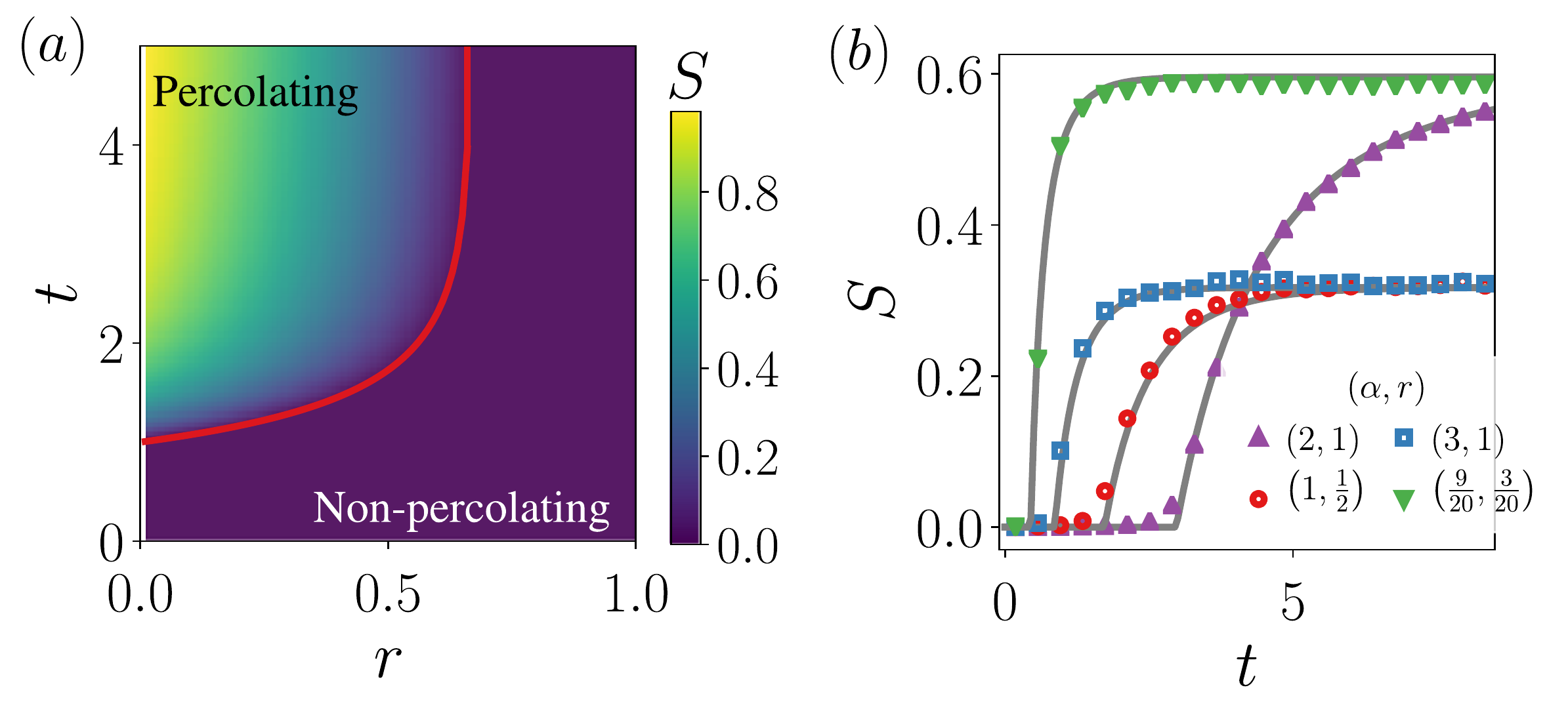}
\caption{In $(a)$, size of the giant component as a function of time and the reset rate. The growth rate is $\alpha = 1$. Separating the percolating and non-percolating phases, there is the critical line $t_c(\alpha, r)$, from Eq.~\eqref{eq:crit_point}. In $(b)$, temporal evolution of the size of the giant component for different values of $\alpha$ and $r$. Solid lines come from the analytical approximations Eqs.~\eqref{eq:SizeGC} and \eqref{eq:RecRel}, markers come from simulations. In all curves, we have considered $N = 20000$ and averages over $10$ realizations.}
\label{fig:fig3}
\end{figure}

We graphically investigate the time-dependent percolation framework developed above, in Fig.~\ref{fig:fig3}$(a)$. We show the size giant component $S$ as a function of time and of the resetting parameter, for fixed $\alpha = 1$. We see that as resetting becomes more and more probable, the giant component needs more time to emerge and its stationary value tends to be smaller. There is a value of resetting rate, $2 \alpha / 3$, above which the giant component never emerges, no matter how long we wait. That is, the minimum average degree (at stationarity) for a network to display a percolating structure is $3/4$. This is somehow counterintuitive at first sight, because we would expect, on average, at least one link per node in order to find a connected subgraph traversing a large fraction of the network. Indeed, for the Erd\H{o}s-R\'{e}nyi model the giant component is found only if $\langle k \rangle > 1$. This conundrum is solved by looking at higher moments of the degree distribution: if $\alpha$ is large enough, or $r$ is small enough ---they need to satisfy $\alpha / r > 3/2$--- the fluctuations from the mean value become relevant and create nodes with large enough degrees, so the giant component can emerge even if, on average, nodes have less than one neighbor. It is not difficult to verify that the Moloy-Reed criterion~\cite{molloy1995critical} for the existence of the giant component precisely leads to $\langle k \rangle^{\text{st}} > 3/4$.
 
To further evince the non-trivial impact of the growth and resetting rate on the percolation transition, in Fig.~\ref{fig:fig3}$(b)$ we show the time evolution of the size of the giant component for several pairs $(\alpha, r)$. We see that if their ratio is the same, the curves tend to the same stationary value, as expected from the long-time limit of the degree distribution and the mean degree. However, the absolute values of $(\alpha, r)$ do have a strong impact on the critical point and on the time scale to reach the stationary value, as it can be appreciated in the ordering of the curves. We see, moreover, that theory and simulations match very well in all the cases. 

\section{First-passage statistics}

\begin{figure}[!t]
  \centering
  \includegraphics[width=0.48\columnwidth]{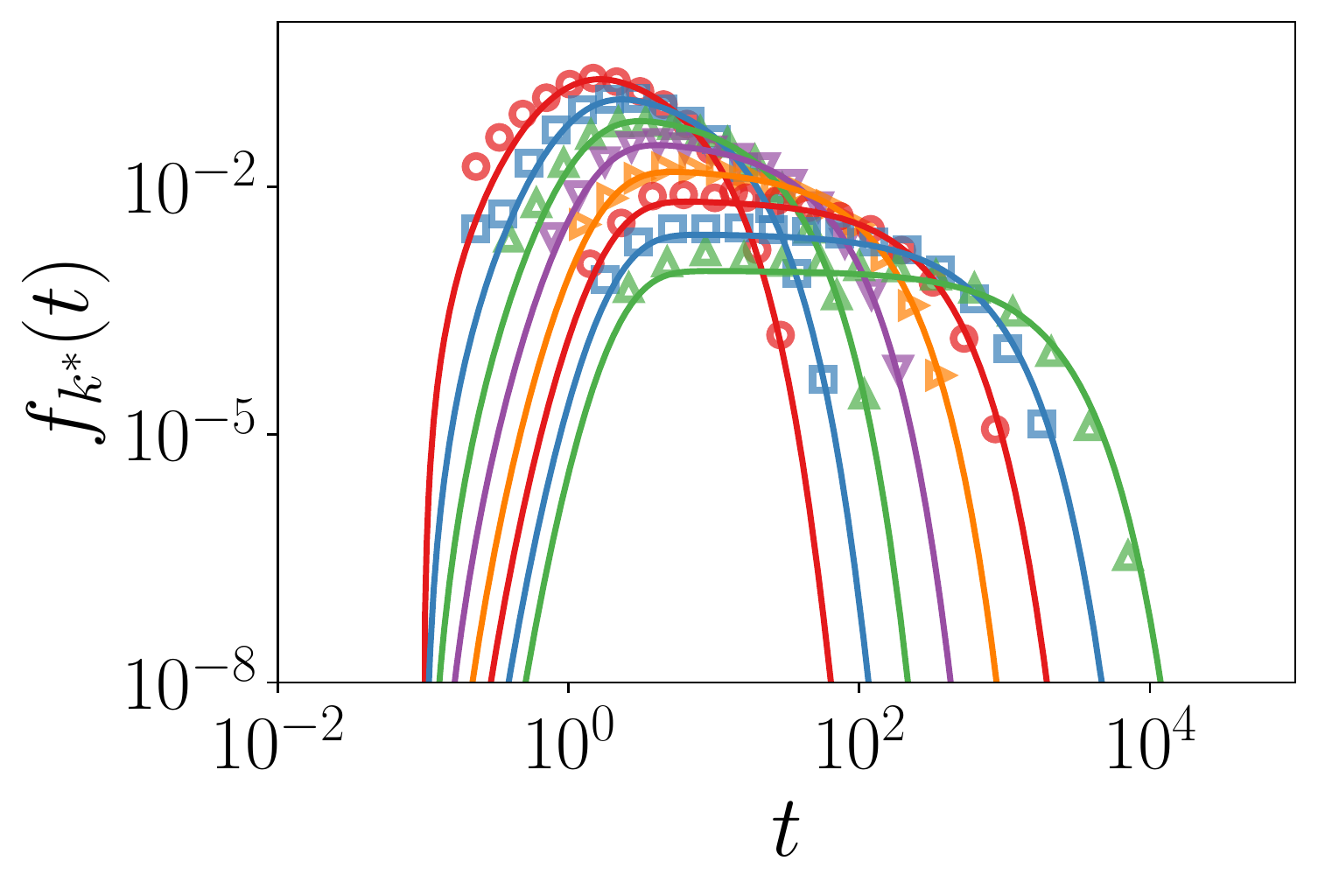}
  \includegraphics[width=0.48\columnwidth]{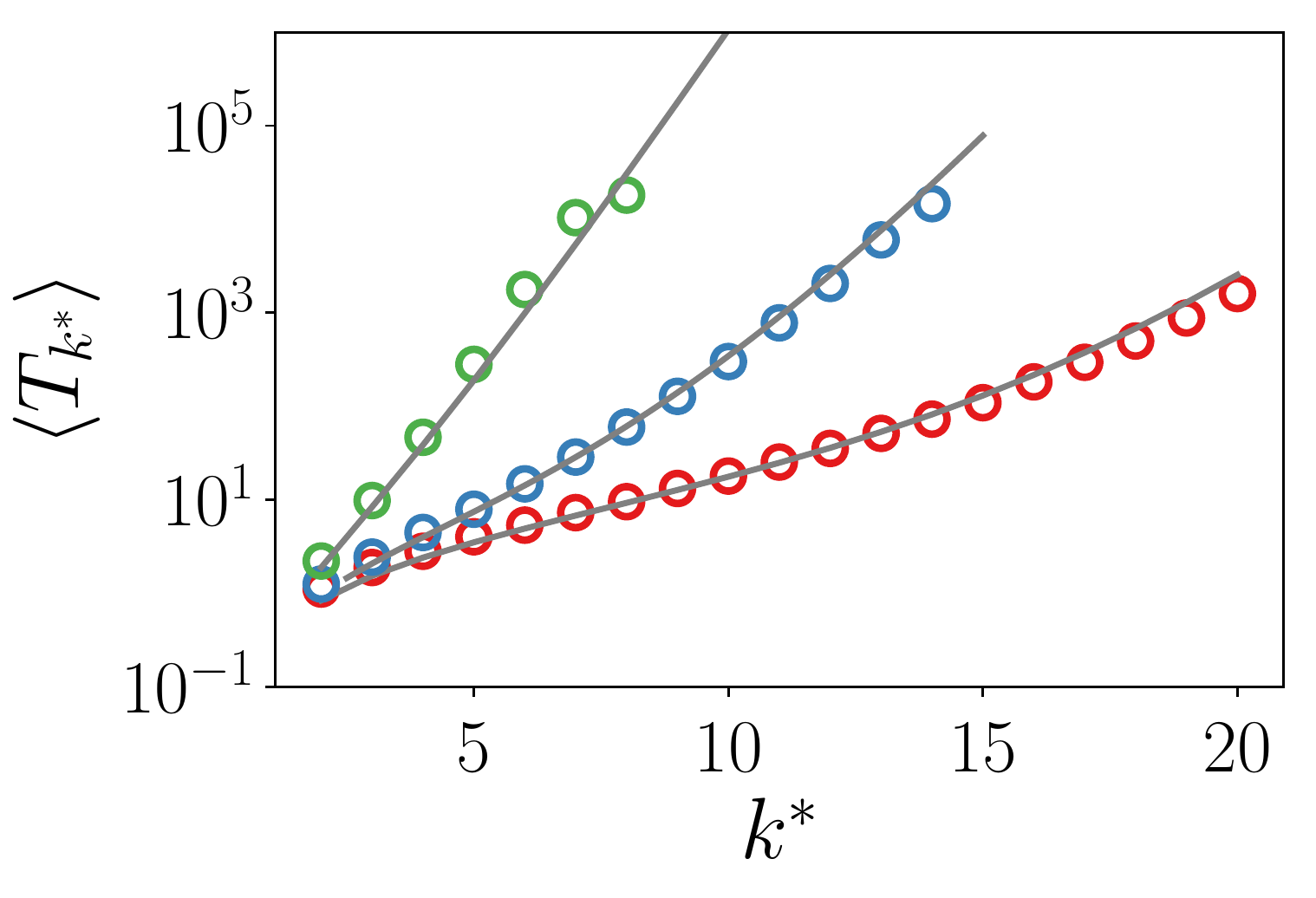}
\caption{In $(a)$, first-passage time distribution to target degrees $k^* = 4, 5, \ldots, 11$. Solid lines correspond to $f_{k^*}$ computed from the numerical integration of Eqs.~\eqref{eq:survprob1}--\eqref{eq:survprob3}, while markers come from simulations. The growth and reset parameters $\alpha = 2$ and $r = 0.5$. In $(b)$, mean first-passage time computed from the analytical expression obtained in the Appendix (solid lines) and from simulations (markers). From left to right, curves corresponding to $(\alpha, r) = (2,2.5)$, $(2,0.5)$ and $(2,0.2)$. In all simulations, we use networks with $N=100$ nodes.}
\label{fig:fig4}
\end{figure}

A final point that we address is the first-passage properties of this coupled multiparticle system. Let $q_k(t)$ be the probability that at time $t$ a randomly selected node has degree $k$ without having arrived at degree $k^* > k$, and let $a_{k^*}(t)$ be the probability that at time $t$ a randomly selected node has arrived at degree $k^*$. Mathematically, this is equivalent to setting an absorbing boundary at $k^*$, so there is no outflow probability from that state. The master equations for these quantities are
\begin{align}
    \diff{q_k} = & \alpha q_{k-1} -  \alpha q_{k} - r q_{k} + r(k+1)q_{k+1} - r k q_{k} + \nonumber \\ & +  \delta_{k,0} r (1 - a_{k^*}) \quad \text{for $0 \leq k \leq k^*-2$}. \label{eq:survprob1} \\
    \diff{q_k} = & \alpha q_{k-1} -  \alpha q_{k} - r q_{k} - r k q_{k} \quad \text{for $k = k^*-1$}. \label{eq:survprob2} \\
    \diff{a_k} = & \alpha q_{k-1} \quad \text{for $k = k^*$}. \label{eq:survprob3}
\end{align}
The initial conditions are $q_k(0) = \delta_{k,0}$. Clearly, the first-passage time probability to the target degree $k^*$ is given by $f_{k^*}(t) = \alpha q_{k^*-1} $, with the mean first-passage time being then $\langle t_{k^*} \rangle = \int \diffint t\, t f_{k^*}$. We show in Fig.~\ref{fig:fig4}$(a)$ that the first-passage distribution obtained from the numerical solution of Eqs.~\eqref{eq:survprob1}--\eqref{eq:survprob3} match well with the results from simulations. 

It is a non-trivial task to obtain the solution of the above set of equations. In order to proceed and find some analytical results regarding the first-passage statistics, we apply some approximations whose validity will be later checked. In particular, we transform the set of discrete master equations~\eqref{eq:survprob1}--\eqref{eq:survprob3} into a single Fokker-Planck equation, approximating the degree as a continuous variable (see Appendix for the derivation). If we define $q(k,t)$ as the continuous version of $q_k(t)$, that is, $q(k,t)$ is a probability density function that gives the probability of finding a randomly chosen node with degree $k \in \left[k, k+\diffint{k} \right]$ at time $t \in \left[ t, t + \diffint{t} \right]$ that has not yet arrived at $k^*$, we obtain
\begin{equation}
    \label{eq:FP_main}
    \pdiff{q}{t} = - \pdiff{}{k} \left[ v(k) q(k,t) \right] + \ppdiff{}{k}  \left[ D(k) q(k,t) \right] - r q(k,t) + r \delta(k),
\end{equation}
with the drift coefficient $v(k)=(\alpha - rk )$ and the diffusion coefficient $ D(k) = (rk + \alpha)/2 $. The boundary conditions are $q(k^*,t) = 0$ and $\partial_k q(k,t) |_{k=0} = 0$, and the initial condition is $q(k,0) = \delta(k)$. The survival probability is defined such that
\begin{equation}
    S(k^*, t) = \int_0^{k^*} \, \diffint{k} \, q(k,t),
\end{equation}
that relates to the mean first-passage time as
\begin{equation}
    \langle T_{k^*} \rangle = - \int_0^{\infty} \diffint t\, t \pdiff{S(k^*,t)}{t} = \Tilde{S}(k^*, s=0),
\end{equation}
where $\Tilde{S}(k^*, s) = \int_0^{\infty} \diffint{t} e^{-st} S(k^*,t)$ is the Laplace transform of the survival probability.

In the Appendix, we obtain a solution for the $\Tilde{S}$ in terms of special functions. However, a direct inspection of Equation~\eqref{eq:FP_main} already provides information that would hardly be obtained from the discrete master equations or grasped by analyzing directly the analytical solution. Given the growth and reset parameters, we see that the drift term vanishes at $k_{\text{ip}} = \alpha / r$. Those nodes with degrees $k < k_{\text{ip}}$ experience a positive drift toward $k_{\text{ip}}$, while, on the contrary, those with degrees $k > k_{\text{ip}}$ suffer a negative drift toward $k_{\text{ip}}$. Consequently, we anticipate different behaviors depending on whether the target degree $k^*$ is larger or smaller than  $k_{\text{ip}}$. In particular, in the regime $k^* < k_{\text{ip}}$, the creation of links dominates over the resetting of nodes, and the arrival to the target degree will be relatively fast. In the regime $k^* > k_{\text{ip}}$, however, the resetting dominates and the system does not create connections quick enough in order for a node to easily reach the target degree. Thus, we expect a considerable increase in the mean first-passage time. In fact, because the drift is degree-dependent, we expect that the larger the difference $k^* - k_{\text{ip}} > 0$, the more difficult will be to arrive to the target degree and the longer it will take. Likewise, the larger the difference $k_{\text{ip}} - k^* > 0$, the faster the target will be attained, hence observing a decrease in the MFPT. Note, however, that we should not wait for a non-trivial combination of $(\alpha, r)$ that minimizes the mean first-passage time, because our system is now bounded and the non-resetting transitions are unidirectional. Hence, it is natural to suppose that $\langle T_{k^*} \rangle$ shows a concave-to-convex inflection point at $k^* = k_{\text{ip}}$ if $\alpha > r$. When $\alpha < r$ the inflection point will disappear and the mean first-passage time will be purely convex over all traget degrees $k^*$. In Fig.~\ref{fig:fig4}$(b)$, we verify that the analytical approximation for the MFPT reproduces well the results from simulations. Indeed, we find an overall good agreement, especially in the concave region. When we enter in the convex regime, where resetting dominates, we observe that some slight differences between the prediction and simulations arise, that
can be associated with the failure of the mean-field approach in capturing the full complexity of the growth and reset network process.

\section{Conclusions}

In summary, we have approached a central problem in network science by reformulating it as a stochastic resetting process. In doing so, we have provided insights that are relevant to both areas of research. On the one hand, network growth under node removal advances our knowledge of stochastic resetting, providing an analytically amenable model characterized by many-body variable interactions. In it, the particles' state evolution is non-trivially coupled in the degree space: growth events can be seen as a two-particle coupling, with a simultaneous increase of one degree unit; reset events can be seen as a $(k+1)$-particle coupling, where one node in state $k$ resets and $k$ other nodes lose one degree unit. On the other hand, we have studied several out-of-equilibrium properties of network growth with node removal, as it is customary in stochastic resetting problems.

In particular, we have obtained an exact expression for the time-dependent degree distribution, which has allowed us to elucidate network-wide properties such as the existence of a connected component occupying a macroscopic portion of the network. In addition, we have studied first-passage statistics, finding that our system does not display a minimum in the mean first-passage time from the origin to a target state $ k^* > 0 $, typical of many processes with resetting~\cite{evans2020stochastic}. Instead, we find a monotonously increasing function that, however, can present an inflection point depending on the growth and resetting rates. 

There are several directions for further research. One deals with the inclusion of more realistic dynamical rules both in the network formation and in the resetting process. This would partially change the mathematical form of the master equations and the challenge would be related to solving more complicated equations that could lead to new physical insights. On these lines, it would be interesting to test the performance of different growth and reset mechanisms in reproducing the evolution of empirical networks. Another research avenue deals with a more complete characterization of the very same model presented in this article. For example, when computing first-passage statistics we could be interested in knowing the first-passage time of the $i$th fastest node arriving to a target. The cluster dynamics would be very interesting to be explored as well, as network growth with node removal can be seen as a highly non-trivial aggregation-fragmentation process for which writing a Smoluchowski-like equation and grasping analytical insights remain as a considerable technical challenge. Finally, we acknowledge that correlated transitions like the ones presented here are still an under-researched characteristic in stochastic resetting models. Therefore, devising new mathematically tractable models that incorporate this feature and that lend themselves to be eventually compared to empirical data is a necessary step toward a more complete, solid and useful theory of stochastic resetting. 

\section*{Acknowledgments}
The author thanks M. De Domenico and D. Kiziridis for their comments on the manuscript.

\setcounter{equation}{0}
\renewcommand{\theequation}{A.\arabic{equation}}

\section*{Appendix: Continuous approach}

To gain insight into the first-passage problem, we proceed by applying some approximations, namely, we are treating the discrete degrees $k$ as a continuous variable. We will see that even for small target degrees $k^*$, hence a small effective interval, the approximation works well. 

Our starting point are the master equations~\eqref{eq:survprob1}--\eqref{eq:survprob3}. Let $q(k,t)$ be continuous version of $q_k(t)$, that is, $q(k,t)$ is a probability density function that gives us the probability of finding a node with degree $k \in \left[k, k+\diffint{k} \right]$ at time $t \in \left[ t, t + \diffint{t} \right]$. We notice that we can write
\begin{equation}
    \pdiff{q}{t} = \Omega^+(k - \delta k) q(k - \delta k, t) + \Omega^-(k + \delta k) q(k + \delta k, t) -  \left[\Omega^+(k) + \Omega^-(k) + \rho(k)\right] q(k, t),
\end{equation}
where $\Omega^+(k) = \alpha$ is the growing rate, while $\Omega^-(k) = rk$ and $\rho(k) = r$ correspond to the resetting rates, the former due to a neighbor deletion and the latter due to the deletion of the node itself. For convenience, we leave $\delta k$ undetermined during the calculations and set $\delta k = 1$ at the end. Expanding both the rates and the pdf up to order $(\delta k)^2$, we arrive at
\begin{equation}
    \pdiff{q}{t} = - \pdiff{}{k} \left[ v(k) q(k,t) \right] + \ppdiff{}{k}  \left[ D(k) q(k,t) \right] - \rho(k) q(k,t),
\end{equation}
where the drift and diffusion terms are, respectively, $v(k) = \delta k (\Omega^+(k) - \Omega^-(k))$ and $D(k) = \delta k^2 (\Omega^+(k) + \Omega^-(k))/2$. We have not yet incorporated the information given in master equations for the boundary degrees $k=0$ and $k=k^*$. The former can be manually introduced as a delta-like source of probability at the resetting point, $k_{\text{reset}}$,while the former takes the form of an absorbing boundary condition. Putting all the pieces together, we are dealt with

\begin{equation}
    \pdiff{q}{t} = - \delta k \pdiff{}{k} \left[ (\alpha - rk) q(k,t) \right] + \frac{\delta k^2}{2} \ppdiff{}{k} \left[ (rk + \alpha) q(k,t) \right] - r q(k,t) + r \delta(k-k_{\text{reset}}),
\end{equation}
with $q(k^*,t) = 0$, $\partial_k q(k,t) |_{k=0} = 0$ and $q(k,0) = \delta(k-k_0)$.

In this continuous-degree approach, the survival probability reads $S(k^*, k_0,t) = \int_0^{k^*} \, \diffint{k} \, q(k,t)$, where $k_0$ is the initial degree. Treating $k_0$ as a variable, the survival probability satisfies the backward equation
\begin{equation}
    \pdiff{S}{t} = (\alpha -rk) \pdiff{S}{k_0} + \frac{\delta k^2}{2} (rk + \alpha) \ppdiff{S}{k_0}  - r S(k^*,k_0,t) + r S(k^*,k_{\text{reset}},t).
\end{equation}
with $S(k^*,k^*, t) = 0$ and $\partial_{k_0} S(k^*,k_0, t)|_{k_0=0} = 0$. Introducing the Laplace transform $\Tilde{S}(k^*,k_0, s) = \int_0^{\infty} \diffint{t} e^{-st} S(k^*,k_0,t)$, we obtain
\begin{equation}
    (\alpha - rk) \pdiff{\Tilde{S}}{k_0} + \frac{rk + \alpha}{2} \ppdiff{\Tilde{S}}{k_0}  - (r + s) \Tilde{S}(k^*,k_0,s) = - 1 - r \Tilde{S}(k^*,k_{\text{reset}},t).
\end{equation}
The general solution to this equation is 
\begin{equation}
    \Tilde{S}(k^*,k_0,s) = \frac{r \Tilde{S}(k^*, k_{\text{reset}},t) + 1}{s+r} + c_1 \mathcal{U}(1,k_0) + c_2 \mathcal{L}(0,k_0),
\end{equation}
where to ease the notation we have introduced the functions $\mathcal{U}(x,y) = U(x + r/\alpha, x - 1 + 4 \alpha/ r, 2y + 2\alpha/r)$, being $U(\cdot, \cdot, \cdot)$ the confluent hypergeometric function of the second kind, and $\mathcal{L}(x,y) = L_{-(x+1) - s/r}^{x-1+ 4\alpha/r} (2y + \alpha/r)$, being $L_{a}^{b} (\cdot)$ the generalized Laguerre polynomial. The derivative is
\begin{equation}
    \pdiff{\Tilde{S}}{k_0} = -\frac{2}{r} (r+s) \, c_1 \,  \mathcal{U}(2,k_0) -2 \, c_2 \, \mathcal{L}(1,k_0).
\end{equation}
Applying the boundary conditions, after some easy algebra we obtain
\begin{align}
    c_1 & = \frac{r \Tilde{S}(k^*, k_{\text{reset}},t) + 1}{s+r} \frac{B_2}{A_2 B_1 - A_1 B_2}  \\
    c_2 & = \frac{r \Tilde{S}(k^*, k_{\text{reset}},t) + 1}{s+r} \frac{A_2}{A_1 B_2 - A_2 B_1},
\end{align}
with
\begin{align}
    A_1 & = \mathcal{U}(1, k^*),  \\
    B_1 & = \mathcal{L}(0, k^*),  \\
    A_2 & = -\frac{2}{r} (r+s) \mathcal{U}(2, 0)  ,  \\
    B_2 & = -2 \mathcal{L}(1, 0).
\end{align}
Note that all these quantities are constant with respect to the ``spatial'' variable $k_0$ but of course do depend explicitly on the growth and reset rates, $\alpha$ and $r$, as well as the target degree $k^*$ and the time in Laplace domain $s$. Now setting $k_0 = k_{\text{reset}} = 0$, we obtain the Laplace transform of the probability that a randomly chosen node starts with no connections and arrives for first time to reach degree $k^*$. Taking the limit $s \to 0$, we have an analytical expression for the Laplace transform of the survival probability, from which we can obtain the mean first-passage time of a randomly chosen node of an initially empty network to reach degree $k^*$. 

Note that if we relax the condition $k_{\text{reset}} = 0$,  the value and relative ordering of $k_{\text{reset}}$, $k^*$, and $k_{\text{ip}}$ (the degree value at which the drift vanishes, see main text) will impact in the behavior of the first-passage distribution and its moments. Despite being an interesting theoretical exercise, these cases do not bring new phenomenological insights with respect to the case $k_{\text{reset}} = 0$. For this reason, a careful report of all the combinations falls outside the scope of the present article, and we stick to the case of nodes losing all their links in the resetting events.

\section*{References}
\bibliographystyle{iopart-num}
\bibliography{biblio}
\end{document}